\documentclass[epj]{svjour}
\usepackage{graphics}
\begin{document}
\title{QGP and Modified Jet Fragmentation}


\author{Xin-Nian Wang 
}                     
\offprints{}          
\institute{Nuclear Science Division, MS 70R319,
Lawrence Berkeley National Laboratory, Berkeley, CA 94720}
\date{Received: date / Revised version: date}
%
\abstract{ Recent progresses in the study of jet modification in
hot medium and their consequences in high-energy heavy-ion
collisions are reviewed. In particular, I will discuss energy loss
for propagating heavy quarks and the resulting modified
fragmentation function. Medium modification of the parton
fragmentation function due to quark recombination are formulated
within finite temperature field theory and their implication on
the search for deconfined quark-gluon plasma is also discussed.
} 

\maketitle
\section{Introduction}
\label{intro}

Jet quenching in high-energy nuclear collisions has been proposed
as a good probe of the hot and dense medium \cite{GP,WG} formed in
ultra-relativistic heavy-ion collisions. The quenching of an
energetic parton is caused by multiple scattering and induced
parton energy loss during its propagation through the hot QCD
medium. Recent theoretical estimates \cite{GW1,BDMPS,Zh,GLV,Wie}
all show that the effective parton energy loss is proportional to
the gluon density of the medium. Therefore measurements of the
parton energy loss will enable one to extract the initial gluon
density of the produced hot medium. Because of color confinement
in the vacuum, one can never separate hadrons fragmenting from the
leading parton and particles materializing from the radiated
gluons. The total energy in the conventionally defined jet cone in
principle should not change due to induced radiation, assuming
that most of the energy carried by radiative gluons remains inside
the jet cone \cite{baier99}. Additional rescattering of the
emitted gluon with the medium could broaden the jet cone
significantly, thus reducing the energy in a fixed cone. However,
fluctuation of the underlying background in high-energy heavy-ion
collisions makes it very difficult, if not impossible, to
determine the energy of a jet on an event-by-event base with
sufficient precision to discern a finite energy loss of the order
of 10 GeV. Since high-$p_T$ hadrons in hadronic and nuclear
collisions come from fragmentation of high-$p_T$ jets, energy loss
naturally leads to suppression of high-$p_T$ hadron spectra
\cite{WG}.

Since parton energy loss effectively slows down the leading parton in
a jet, a direct manifestation of jet quenching is the modification of the
jet fragmentation function, $D_{a\rightarrow h}(z,\mu^2)$, which can
be measured directly in events in which one can identify the jet via a
companion particle like a direct photon \cite{wh} or a triggered high
$p_T$ hadron. This modification can be directly translated into the
energy loss of the leading parton. Since inclusive hadron
spectra are a convolution of the jet production cross section and the
jet fragmentation function in pQCD, the suppression of
inclusive high-$p_T$ hadron spectra is a direct consequence
of the medium modification of the jet fragmentation function
caused by parton energy loss.

Strong suppression of high transverse momentum hadron spectra is
indeed observed by experiments \cite{Phenix,Star} at the
Relativistic Heavy-Ion Collider (RHIC) at the Brookhaven National
Laboratory (BNL), indicating large parton energy loss in a medium
with large initial gluon density. Shown in Fig.~\ref{fig1} are
the nuclear modification factors $R_{AA}(p_T)$ for single hadron
spectra as a function of the number of participant nucleons. The
theoretical results are obtained from a LO pQCD parton model
calculation \cite{Wang:2003mm} incorporating modified parton
fragmentation functions due to parton energy loss,
\begin{equation}
\langle \Delta E \rangle \approx \pi C_aC_A\alpha_s^3
\int_{\tau_0}^{R_A} d\tau \rho(\tau) (\tau-\tau_0)\ln\frac{2E}{\tau\mu^2}.
\label{effloss}
\end{equation}
The initial gluon density in the most central $Au+Au$ collisions
at $\sqrt{s}=200$ GeV was fixed by fitting the data. The
centrality dependence shown is the consequence of the above parton
energy loss, assuming the initial gluon density is proportional to
the measured hadron multiplicity which in turn is approximately
proportional to the number of participant nucleons. Such a
calculation also has a definite prediction of the energy
dependence of the hadron spectra suppression \cite{Wang:2003aw}
that agrees well with the current collection of data at different
energies \cite{d'Enterria:2005cs}.

\begin{figure}
\resizebox{0.4\textwidth}{!}{%
  \includegraphics{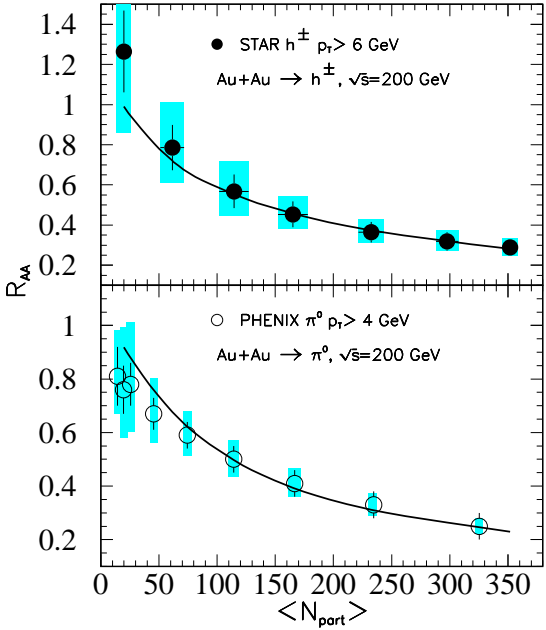}}
\caption{The centrality dependence of the measured single inclusive
hadron suppression \protect\cite{phenix-r2,star-r2} at high-$p_T$ as
compared to theoretical calculation with parton energy loss}
\label{fig1}
\end{figure}

Since the parton energy loss [Eq.~(\ref{effloss})] depends on the
path length of the jet propagation which in turn depends on the
azimuthal angle with respect to the reaction plane in non-central
collisions, the parton energy loss and modified fragmentation
functions naturally lead to an azimuthal angle dependence of the
hadron spectra suppression \cite{wangv2,gvwv2}. Indeed, the
azimuthal angle distributions of high $p_T$ hadrons were found to
display large anisotropy with respect to the reaction planes
\cite{star-jetv2} of non-central $Au+Au$ collisions. As shown in
Fig.~\ref{fig2}, the observed azimuthal anisotropy, characterized
by the second coefficient of the Fourier transformation $v_2$, can
also be described well by the same pQCD parton model calculation.

\begin{figure}
\resizebox{0.45\textwidth}{!}{%
  \includegraphics{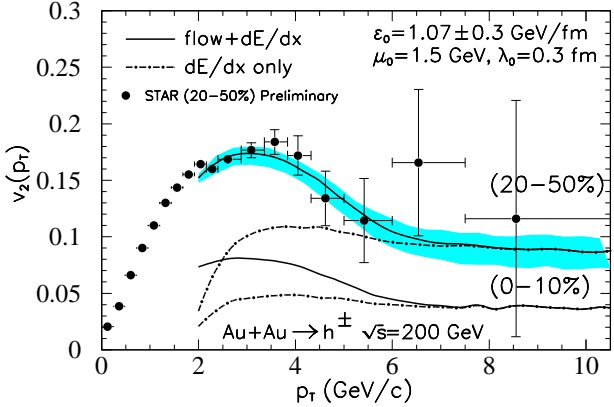}}
\caption{Azimuthal anisotropy in $Au+Au$ collisions
as compared to the STAR\protect\cite{Snellings:2003mh} 4-particle
cumulant result.}
\label{fig-highptv2}
\end{figure}

The most striking measurement that is a direct manifest of jet quenching
is the observed disappearance of the back-side high-$p_T$ two-hadron
correlation in azimuthal angle \cite{star-jet}, which is characteristic
of high-$p_T$ back-to-back jets in $p+p$ collisions. Shown
in Fig.~\ref{fig3}, are back-side two high $p_T$ hadron correlations
in $Au+Au$ collisions with different centralities as compared to
the same correlation in $p+p$ collisions. Though the back-side
correlations in peripheral $Au+Au$ collisions remain the same as
in $p+p$ collisions, the peak gradually decreases and finally disappears
in the most central $Au+Au$ collisions. The curves are again from
the same LO pQCD parton model calculation with parton energy loss.
It describes both the magnitude of the suppression and the
centrality dependence quite well.

Combining the above measurements of three different effects of
parton energy loss and compare with the measured jet quenching in
deeply inelastic $e+A$ collisions, one can conclude that the
initial gluon density reached in central $Au+Au$ collisions at
$\sqrt{s}=200$ GeV is about 30 times higher than in a cold $Au$
nuclei \cite{Wang:2003mm,ww02}, assuming the theoretical result
that parton energy loss is proportional to the gluon density of
the medium.

\begin{figure}
\resizebox{0.4\textwidth}{!}{%
  \includegraphics{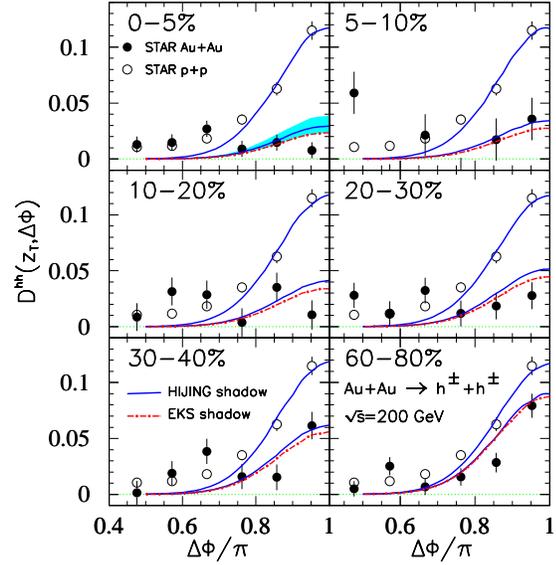}}
\caption{Back-to-back correlations for charged hadrons
with $p^{\rm trig}_T>p_T>2$ GeV/$c$,
$p^{\rm trig}_T=4-6$ GeV/$c$ and $|y|<0.7$ in $Au+Au$ (lower curves)
and $p+p$ (upper curves)
collisions as compared to the STAR\protect\cite{star-jet} data.
\label{fig-b2b}}
\end{figure}

In this talk, I will review recent progresses we have made in the
study of modified jet fragmentation functions in medium and their
applications to high $p_T$ hadron spectra and correlations in
heavy-ion collisions. In particular, I will review heavy quark
energy loss and modified fragmentation functions. Its unique
features due to the heavy quark mass can help us to characterize
the partonic nature of the observed jet quenching. I will also
discuss dihadron fragmentation functions since it addresses hadron
correlation within a jet and how it will also be modified by the
parton energy loss. Finally, I will discuss the formulation of
modified fragmentation functions due to quark recombination during
the the hadronization of the jet in a thermal medium. The
implication on the search for deconfined quark-gluon plasma will
also be discussed.

\section{Modified Heavy Quark Fragmentation Function}
\label{sec:3}

In the study of parton energy loss, formation time for the radiated
gluon plays an essential role. Because of the LPM interference,
gluons with formation time longer than the length of
a finite medium or the mean-free path in an infinitely large
medium will be suppressed. Such formation time is only
relative to the propagation of the leading parton.
Therefore, gluon radiation from a heavy quark is normally shorter than
that from a light quark because of the smaller velocity of the
heavy quark. For radiated gluons with transverse momentum $\ell_\perp$
and $z$ fractional momentum, the formation time is \cite{Zhang:2003wk}
\begin{equation}
\tau_f =\frac{2z(1-z)E}{\ell_T^2+(1-z)^2M^2},
\end{equation}
where $M$ is the quark mass. Therefore, one should expect the LPM
effect to be significantly reduced for intermediate
energy heavy quarks. In addition, the heavy quark mass also
suppresses gluon radiation amplitude at small angles \cite{DK}
relative to that off a light quark. Both mass
effects will lead to a reduced heavy quark energy loss compared to
that of a light quark. The most significant consequence of
the reduced formation time due to heavy quark mass is the change
of length dependence of the quark energy loss. The non-Abelian
LPM effect due to suppression of gluon radiation with long formation
time leads to a quadratic length dependence of the total energy loss.
For a slow heavy quark, however, the dependence will become linear
because of the short formation time and absence of the LPM interference.
Shown in Fig.~\ref{fig1} are the numerical results of the
nuclear size $R_A$ dependence of charm quark fractional energy loss
in DIS off a cold nucleus, rescaled
by $\widetilde{C}(Q^2)C_A\alpha_s^2(Q^2)/N_C$, for different
values of $x_B$ and $Q^2$. One can clearly see that the $R_A$ dependence
is quadratic for large values of $Q^2$ or small $x_B$ (large initial
quark energy) when the mass of the quark is negligible.
The dependence becomes almost linear for small $Q^2$ or large $x_B$.
The charm quark mass is set at $M=1.5$ GeV in the numerical calculation.

\begin{figure}
\resizebox{0.45\textwidth}{!}{%
  \includegraphics{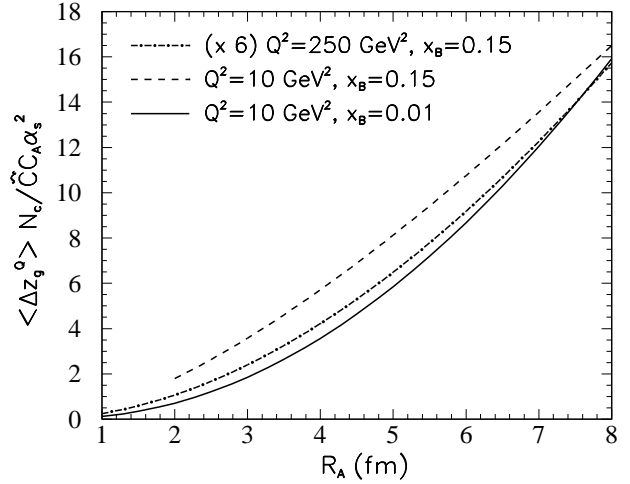}}
\caption{
The nuclear size, $R_A$, dependence of charm quark energy
loss for different values of $Q^2$ and $x_B$.}
\label{fig1}       
\end{figure}

\begin{figure}
\resizebox{0.5\textwidth}{!}{%
  \includegraphics{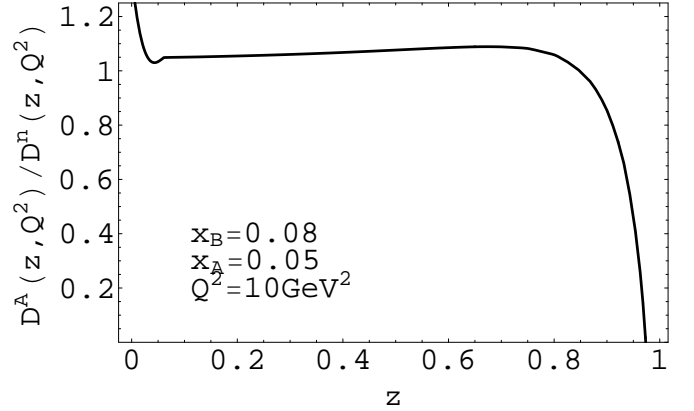}}
\caption{
Modification factor for the charm quark fragmentation
function in a nucleus. The value $x_A=0.05$ corresponds to a
nucleus with a radius $R_A=4.25$ fm.}
\label{fig2}       
\end{figure}

One can similarly calculate the nuclear modification of the heavy
quark fragmentation as in the case of a light quark \cite{ww02}.
Shown in Fig.~\ref{fig2} is the ratio of the modified charm
quark fragmentation into $D$ mesons to the vacuum fragmentation
function. One can see that the modification due to the parton
rescattering and induced gluon radiation in a nucleus for heavy
quarks is quite different from light quarks \cite{ww02,GW}.
This is mainly caused by the form of the heavy quark fragmentation
function in vacuum which peaks at large $z$. Because
of the multiple scattering and induced gluon radiation,
the position of the peak of the modified fragmentation function
is effectively shifted to a smaller value of $z$.
As a consequence, the heavy quark fragmentation function remains
unchanged, or even slightly enhanced for a large range of
fractional momentum $z$.
The modification only becomes significant and the
fragmentation function is suppressed at large $z$ above the
position of the peak. This is in sharp contrast
to the case of modified light quark fragmentation functions
which deviate from the vacuum form in a very large range of $z$.

\section{Non-Abelian feature of jet quenching}
\label{sec:3}

Another non-Abelian feature of the parton energy loss is its
dependence on the color representation of the propagating parton.
The energy loss for a gluon is 9/4 times larger than a quark. One
can investigate the consequences of this non-Abelian feature in
the flavor dependence of the high-$p_T$ hadron suppression
\cite{Wang:1998bh}. In the mean time, we can also study the effect
of the non-Abelian parton energy loss on the energy dependence of
the inclusive hadron spectra suppression \cite{Wang:2004tt}. One
can exploit the well-known feature of the initial parton
distributions in nucleons (or nuclei) that quarks dominate at
large fractional momentum ($x$) while gluons dominate at small
$x$. Jet or large $p_T$ hadron production as a result of hard
scatterings of initial partons will be dominated by quarks at
large $x_T=2p_T/\sqrt{s}$ and by gluons at small $x_T$. Since
gluons lose 9/4 more energy than quarks, the energy dependence of
the large (and fixed) $p_T$ hadron spectra suppression due to
parton energy loss should reflect the transition from
quark-dominated jet production at low energy to gluon-dominated
jet production at high energy. Such a unique energy dependence of
the high-$p_T$ hadron suppression can be tested by combining
$\sqrt{s}=200$ AGeV data with lower energy data or future data
from LHC experiments.

\begin{figure}
\rotatebox{-90}{\resizebox{0.34\textwidth}{!}{%
  \includegraphics{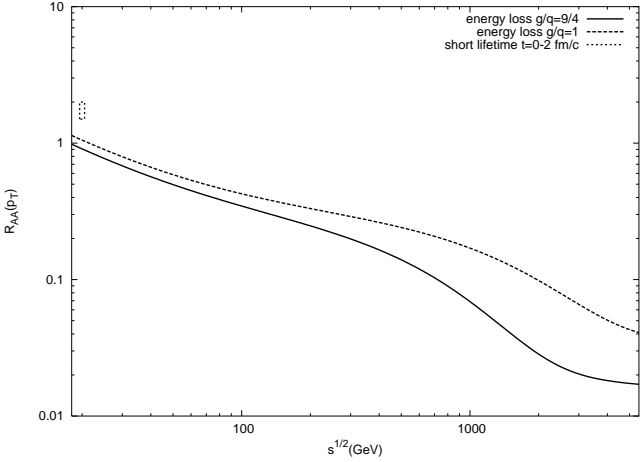} }}
\caption{
Nuclear modification factor $R_{AuAu}$ for neutral pions
as function of collision energy at fixed $p_T=6$ GeV in most central
collisions (with centrality 10\%). Here we compare the QCD energy loss
and a non-QCD one where the energy loss is identical for quarks and gluons.}
\label{fig3}
\end{figure}

To study the sensitivity of hadron spectra suppression to the
non-Abelian parton energy loss, we calculate the single hadron
spectra and the suppression factor with two different parton
energy loss: one for the QCD case where the energy loss for a
gluon is 9/4 times as large as that for a quark, i.e. $\Delta
E_g/\Delta E_q=9/4$; the other is for a non-QCD case where the
energy loss is chosen to be the same for both gluons and quarks.
Similarly, the average number of inelastic scatterings obeys
$\langle \frac{\Delta L}{\lambda}\rangle _g/ \langle \frac{\Delta
L}{\lambda}\rangle _q=9/4$ in the QCD case. For the non-QCD case
we are considering, the above ratio is set to one. In order to
demonstrate the difference between QCD and non-QCD energy loss, we
compute the $R_{AA}$ for neutral pions at fixed $p_T=6$ GeV in
central $Au+Au$ collisions as a function of $\sqrt{s}$ from 20
AGeV to 5500 AGeV. Shown in Fig.\ \ref{fig3} are the calculated
results with both the QCD and non-QCD energy loss. In the
calculation, we fix the initial gluon number density and quarks'
mean-free path to fit the overall hadron suppression in the most
central $Au+Au$ collisions at $\sqrt{s}=200$ GeV. For any other
energy and centralities, we simply assume the initial gluon number
is proportional to the final measured total charged hadron
multiplicity per unit rapidity. One can see that due to the
dominant gluon bremsstrahlung or gluon energy loss at high energy
the $R_{AA}$ for the QCD case is more suppressed than the non-QCD
case where the gluon energy loss is assumed to take an equal role
as the quark.

Another interesting feature with the energy dependence of $R_{AA}$
is the change of slope around $\sqrt{s}=1300$ GeV. The rapid decrease
of $R_{AA}$ at $\sqrt{s}=20-1300$ GeV is mainly due to
increased initial gluon density and also
the change of $p_T$ slope of jet production cross section with $\sqrt{s}$.
As the energy loss increases, more jets produced inside the
overlapped region are completely suppressed. Only those that
are produced within an out-layer in the overlapped region will survive.
This will be like surface emission with a finite depth.
The suppression factor $R_{AA}$ will then be determined by the width
of the out-layer which is just the averaged
mean-free-path $\langle \lambda \rangle$.
As a consequence, $R_{AA}$ will then have much weaker $\sqrt{s}$
dependence. This effect was also seen by calculations in 
Ref.~\cite{Eskola:2004cr}.

\section{Modified dihadron fragmentation function}
\label{sec:4}

In addition to single inclusive hadron spectra from jet
fragmentation, multiple hadron correlations have proven to be a
useful measurement for characterizing the modification of the jet
structure in hot medium. For example, while the back-side
two-hadron correlation is completely suppressed, the same-side
correlation is observed to change little when the transverse
momentum of the secondary hadron is large \cite{star-jet}. Such
same-side correlation is essentially given by dihadron
fragmentation functions. It is important to investigate why such a
dihadron fragmentation function changes little in the kinematic
region of the experiments and whether it will be significantly
modified when the secondary hadron is soft.

One can define the dihadron fragmentation functions in terms of the
overlapping matrix between parton field operators and the final
hadron states, similarly as the single hadron fragmentation functions.
In the light-cone gauge, the dihadron fragmentation function of a
quark are defined as
 \begin{eqnarray}
  D_{q}^{h_1 h_2}(z_{h_1},z_{h_2})
  &=&\frac{z_{h}^4}{2z_{h_1}z_{h_2}}
  \int\frac{d^2 p_{h_{1\perp}}}{2(2\pi)^3}
  \int\frac{d^4p}{(2\pi)^4}
   \nonumber\\
  &\times&\delta\left(z_h - \frac{p_h^+}{p^+}\right)  \int d^4 xe^{-ip\cdot x}
   \nonumber\\
&&\hspace{-1.3in}{\rm Tr}\left\lbrack\frac{\gamma^+}{2 p_{h}^+}
    \sum_{S}\langle 0\left|{\psi}(0)\right|S, p_{h_1},p_{h_2}\rangle
\langle p_{h_2},p_{h_1},S\left|\overline\psi(x)\right|0\rangle\right\rbrack\, ,
 \label{gfrag2}
 \end{eqnarray}
where $z_h=z_{h_1}+z_{h_2}$ and $p_h=p_{h_1}+p_{h_2}$. Like single hadron
fragmentation functions, dihadron fragmentation functions also contain
non-perturbative physics and thus are not calculable in pQCD. However,
one can study their evolution with the energy scale within pQCD. The
DGLAP evolution equations for dihadron fragmentation functions
have been derived recently \cite{majumder1}. For dihadron
fragmentation function of a quark, it has a form
 \begin{eqnarray}
   \frac{\partial D_q^{h_1 h_2}(z_{h_1}, z_{h_2}, Q^2)}{\partial\log Q^2}
 &=& \nonumber \\
 & &  \hspace{-1.0in}\frac{\alpha_s}{2\pi}
 \left[\int_{z_{h_1}+z_{h_2}}^1\frac{dz}{z^2}\gamma_{qq}(z)
   D_q^{h_1 h_2}\left(\frac{z_{h_1}}{z},\frac{z_{h_2}}{z},Q^2\right)\right.
 \nonumber\\
  &&\hspace{-1.5in}+
  \int_{z_{h_1}}^{1-z_{h_2}}\frac{dz}{z(1-z)}{\tilde \gamma}_{qq}(z)
   D_q^{h_1}\left(\frac{z_{h_1}}{z},Q^2\right)
   D_g^{h_2}\left(\frac{z_{h_2}}{1-z},Q^2\right)
 \nonumber\\
   &&\hspace{-1.5in}
   + (1\leftrightarrow 2)+\left.\int_{z_{h_1}+z_{h_2}}^1\frac{dz}{z^2}\gamma_{qg}(z)
   D_g^{h_1 h_2}\left(\frac{z_{h_1}}{z},\frac{z_{h_2}}{z},
   Q^2\right)\right]\, ,
 \label{DGLAP21}
\end{eqnarray}
that is similar to the DGLAP evolution equations for single hadron
jet fragmentation functions. However, there are extra contributions in
the above equation that are proportional to convolution of two
single hadron fragmentation functions. These correspond to independent
fragmentation of both daughter partons after the parton split in the
radiative correction. Here $\gamma_{qq}$ is the normal parton splitting
function and $\tilde \gamma_{qq}$ is the same function without the virtual
corrections. For numerical solution of the DGLAP evolution
equations of the dihadron fragmentation, results from JETSET \cite{jetset}
at a given scale is used. The evolved dihadron fragmentation functions
from the DGLAP equations at higher scales agree with the Monte Carlo
results very well \cite{majumder1}. It will be useful to compare actual
experimental data when they become available.

Since the induced bremsstrahlung in medium is similar to that in
vacuum, medium modification to the jet fragmentation functions
should resemble the radiative corrections in vacuum that lead
to the DGLAP evolution equations. Therefore, it is not surprising
that the medium modification to the dihadron fragmentation functions
has the identical form as the above DGLAP evolution equations \cite{majumder2}.
These medium modifications depend on the same gluon correlation functions
as in the modification to the single hadron fragmentation functions.
Therefore, in the
numerical calculation of the medium modification of dihadron fragmentation
functions, there are no additional parameters involved. The predicted results
are in good agreement with HERMES data \cite{majumder2}. We find that most of
the nuclear modification is manifested in the single hadron fragmentation
functions. Since dihadron fragmentation functions already contain the information
of single hadron fragmentation function, the modification to the remaining
correlated distribution is very small. So the normalized
correlation $D_q^{h_1h_2}(z_1,z_2)/D_q^{h_1}(z_1)D_q^{h_2}(z_2)$ has much
smaller nuclear modification as compared to the single hadron fragmentation
functions. This also explains why the same-side two-hadron correlation
in central heavy-ion collisions remains approximately the same as in $p+p$
collisions while the back-side is completely suppressed \cite{star-jet}.
However, because of trigger bias, one might sample different values of $z_1$
and $z_2$ in $Au+Au$ and $p+p$ collisions. This could lead to apparent
change of dihadron correlation \cite{majumder2}.

\section{Jet fragmentation and quark recombination}
\label{sec:5}

During the propagation and interaction inside a deconfined hot
partonic medium, a fast parton has not only induced gluon
radiation but also induced absorption of the surrounding thermal
gluons. This leads to a stronger energy dependence of the net
energy loss for an intermediate energy parton \cite{ww01}. In
principle, one can consider such processes of detailed balance as
parton recombination and they can continue until the hadronization
of the bulk partonic matter. Eventually, during the hadronization,
partons from the jet can combine with those from the medium to
form final hadrons. Indeed, there exists already some evidence for
the parton recombination in the experimental data on the final
hadron spectra in heavy-ion collisions at RHIC. At intermediate
$p_T=2-4$ GeV/$c$, the suppression of baryons due to jet quenching
is significantly smaller than mesons leading to a baryon to meson
ratio larger than 1, about a factor of 5 increase over the value
in $p+p$ collisions \cite{Adler:2003cb}. On the other hand, the
azimuthal anisotropy of the baryon spectra is higher than that of
meson spectra. Such a flavor dependence of the nuclear
modification of the hadron spectra and their azimuthal anisotropy
is not consistent with a picture of pure parton energy loss. The
most striking revelation of the underlying mechanism comes from
the empirical observation of the scaling behavior between the
azimuthal anisotropy of baryons and mesons \cite{Adams:2003am},
$v_2^M(p_T/2)/2=v_2^B(p_T/3)/3$, inspired by a schematic model of
constituent quark recombination of hadron production.

Many quark recombination models \cite{Hwa:2002tu,Fries:2003vb}
are successful in describing
the observed flavor dependence of the nuclear modification of hadron
spectra at intermediate $p_T$. These models generally have three
different contributions to
the final hadron spectra. They include recombination of the thermal
quarks in the bulk matter into hadrons which dominate low $p_T$
spectra and recombination between thermal quarks and quarks from
high transverse momentum jets that are responsible for intermediate
$p_T$ hadrons. They all assume a thermal distribution for the medium
quarks and employ constituent quark model for the hadron
wavefunctions which determine the recombination probabilities.
However, current models differ in the determination of the constituent quark
distributions from high $p_T$ jets and there exist ambiguities in the
connection between partons from pQCD hard processes and constituent
quarks that form the final hadrons. Perhaps the most consistent
treatment of the problem is the model by Hwa and Yang \cite{Hwa:2002tu}.
In this model, quark recombination processes are traced back to parton
fragmentation processes in vacuum. They consider the initial
produced hard partons that will evolve into a shower of constituent
quarks which then recombine to form the final hadrons in the
parton fragmentation process. The recombination of the shower quarks
of the parton jets with the medium quarks in heavy-ion collisions
can be carried out straightforward given both the shower and medium
quark distributions. Since this is a phenomenological model,
the nuclear modification of the jet shower quark distributions and
their QCD evolution cannot be calculated in the Hwa-Yang model. The model
has to rely on fitting to the experimentally measured hadron spectra
to obtain the corresponding nuclear modified shower quark distributions
for each centrality of heavy-ion collisions and correlation between shower
quarks are completely neglected.

We have made a first attempt to derive the quark recombination
model for jet fragmentation functions from the field theoretical
formulation and the constituent quark model of hadrons \cite{mww}.
Within the constituent quark model, we consider the parton
fragmentation as a two stage process. The initial parton first
evolves into a shower of constituent quarks that subsequently will
combine with each other to form the final hadrons. Since
constituent quarks are non-perturbative objects in QCD similarly
as hadrons, the conversion of hard partons into showers of
constituent quarks is not calculable in pQCD. However, we can
define constituent quark distributions in a jet as overlapping
matrices of the parton field operator and the constituent quark
states, similarly as the definition of the hadron fragmentation
functions.

Given a hadron's (a meson for example) wavefunction in the constituent
quark model,
 \begin{eqnarray}
  | p_h\rangle &=&\int \frac{d^2k_{1\perp}}{2(2\pi)^3}
    \frac{dx_1}{\sqrt{x_1(1-x_1)}}
  \varphi_h(k_{1\perp}, x_1; -k_{1\perp}, 1-x_1) \nonumber \\
  &\times& |k_{1\perp}, x_1; -k_{1\perp}, 1-x_1\rangle ,
\label{pm}
\end{eqnarray}
and neglecting interferences between recombination of quark and
anti-quark pair with different momentum, one can rewrite the
single inclusive meson fragmentation function as a convolution of
the diquark distribution functions and the recombination
probability,
 \begin{eqnarray}
  D_{q}^{h}(z_{h})
  &\equiv&\frac{z_{h}^3}{2}\int\frac{d^4p}{(2\pi)^4}
   \delta\left(z_{h} {-} \frac{p_h^+}{p^+}\right)
   \int d^4 xe^{-ip\cdot x} \nonumber \\
   &\times&{\rm Tr}\left\lbrack\frac{\gamma^+}{2 p_h^+}
    \sum_S\langle 0\left|{\psi}(0)\right| S,p_{h}\rangle
    \langle p_{h},S\left|\overline\psi(x)\right|0\rangle\right\rbrack\, ,
  \nonumber \\
&\approx & C_h \int_0^{z_h}\frac{dz_1}{2}
  R_h(0_\perp,\frac{z_1}{z_M})
  F_{q}^{q_1{\bar q}_2}(z_1, z_h-z_1)  \, , \nonumber \\
\label{qfrag1}
 \end{eqnarray}
where $C_h$ is a constant representing contributions from interference
processes. The recombination probability is determined by the hadrons'
wavefunction,
 \begin{equation}
R_h(k_{1\perp},\frac{z_1}{z_h})
   \equiv
 \left|\varphi_h (k_{1\perp}, \frac{z_1}{z_h}; -k_{1\perp},
   1-\frac{z_1}{z_h})\right|^2 \, ,
 \label{Rm}
 \end{equation}
and the double constituent quark distribution function is
defined as the overlapping matrix between the current quark
operators and the intermediate constituent quark states,
 \begin{eqnarray}
  F_{q}^{q_1{\bar q}_2}(z_1,z_2)
  &=&\frac{z_{h}^4}{2z_1z_2}
  \int^{\Lambda}\hspace{-6pt}\frac{d^2 k_{1\perp}}{2(2\pi)^3}
 \int\hspace{-6pt}\frac{d^4p}{(2\pi)^4} \int \hspace{-6pt}d^4 x
\delta\left(z_h - \frac{p_h^+}{p^+}\right)
  \nonumber\\
   && \hspace{-0.95in}
     e^{-ip\cdot x}
{\rm Tr}\left\lbrack\frac{\gamma^+}{2 p_{h}^+}
    \sum_{S}\langle 0\left|{\psi}(0)\right|S, k_1,k_2\rangle
    \langle k_2,k_1,S\left|\overline\psi(x)
    \right|0\rangle\right\rbrack .\,\, 
 \label{qfrag03}
 \end{eqnarray}
Here, $p_{h}=k_1+k_2$ and $z_h=z_1+z_2$. $\Lambda$ is the cutoff
for the intrinsic transverse momentum of the constituent quarks
inside a hadron, as provided by the hadron wavefunction. The above
definition of diquark distribution function in a fragmenting
parton jet has exactly the same form as the dihadron fragmentation
functions \cite{majumder1}. One can similarly express the parton
fragmentation functions for baryons in terms of triquark
distribution functions. This is similar in spirit to the Hwa-Yang
recombination model. Given a form of the hadrons' wavefunction in
the constituent quark model, one can in principle extract
constituent quark distribution functions from the measured jet
fragmentation functions. Furthermore, within this framework, one
can also derive the DGLAP evolution equations for the diquark
distribution functions,
 \begin{eqnarray}
   &&Q^2\frac{d}{d Q^2}F_q^{q_1 \bar q_2}(z_1, z_2, Q^2)
   =\frac{\alpha_s(Q^2)}{2\pi}\int_{z_1+z_2}^1\frac{dz}{z^2}\times
   \nonumber \\
&&\hspace{-14pt}\left[\gamma_{qq}(z)
   F_q^{q_1 \bar q_2}(\frac{z_1}{z},\frac{z_2}{z},Q^2)
   +\gamma_{qg}(z)
   F_g^{q_1 \bar q_2}(\frac{z_1}{z},\frac{z_2}{z}, Q^2)\right],
 \label{DGLAPq11}\\
   &&Q^2\frac{d }{d Q^2}F_g^{q_1 \bar q_2}(z_1, z_2, Q^2)
   =\frac{\alpha_s(Q^2)}{2\pi}\int_{z_1+z_2}^1\frac{dz}{z^2}\times
    \nonumber \\
    && \hspace{-14pt}\left[\gamma_{gq}(z)
   F_s^{q_1 \bar q_2}(\frac{z_1}{z},\frac{z_2}{z}, Q^2)
   +\gamma_{gg}(z)
   F_g^{q_1 \bar q_2}(\frac{z_1}{z},\frac{z_2}{z},Q^2)\right].
 \label{DGLAPq12}
 \end{eqnarray}
Note that the above equations are a little different from the
DGLAP evolution equations for dihadron fragmentation functions.
There is no contribution from independent fragmentation in the
diquark distribution. This is because the diquark distribution
functions defined in the context of quark recombination are only
for two quarks whose relative transverse momentum is limited by
the wavefunction. Therefore, change of the initial momentum scale
does not lead to variation of phase space available for the
diquark from the independent fragmentation in the final states.
One can show that combining the above evolution equations for the
double constituent quark distribution functions with the
expression of jet fragmentation function in Eq.~(\ref{qfragm02}),
the DGLAP evolution equations for single inclusive hadron
fragmentation functions can be recovered.

The above reformulation
of the jet fragmentation functions does nothing to simplify the
complexity of jet hadronization. However, extending the formalism
to finite temperature, we can automatically derive the contributions
from recombination between shower and thermal quarks in addition
to soft hadron production from recombination of thermal quarks and
leading hadron from recombination of shower quarks. The shower
and thermal quark recombination involves single quark
distribution functions which are related to the diquark
distributions through sum rules. Therefore, one can consistently
describe three different processes within this formalism.
Since the single and diquark distribution functions are
defined at finite temperature which are different from the
corresponding vacuum distributions, one can also consistently take
into account parton energy loss and  detailed balance effect for jet
fragmentation inside a thermal medium.

\section{Modified jet fragmentation due to quark recombination in medium}
\label{sec:5}

One can study the fragmentation
of a parton jet in medium simply by replacing the vacuum expectation
in the $S$ matrix of the processes or the operator definition of the
parton fragmentation functions by the thermal expectation values,
$\langle 0|{\cal O} |0\rangle \rightarrow
\langle\langle{\cal O}\rangle\rangle$,
 \begin{equation}
 \label{expectation}
  \langle\langle{\cal O}\rangle\rangle=\frac{{\rm Tr}[e^{-{\hat H}/T}{\cal O}]}
    {{\rm Tr}\;e^{-{\hat H}/T}}\, ,
 \end{equation}
where, $\hat H$ is the Hamiltonian operator of the system and
$T$ is the temperature. Therefore, the single hadron
fragmentation at finite temperature for a quark is defined as
 \begin{eqnarray}
   {\tilde D_q^h}(z_h, p^+)&=&\frac{z_{h}^3}{2}\int\frac{d^4p}{(2\pi)^4}
   \delta\left(z_{h} {-} \frac{p_{h}^+}{p^+}\right)
   \int d^4 xe^{-ip\cdot x} \nonumber \\
    & &\hspace{-0.3in}{\rm Tr} \left\lbrack\frac{\gamma^+}{2 p_{h}^+}
    \sum_{S}\langle\langle{\psi}(0)|S,p_{h}\rangle
    \langle p_{h},S|\overline\psi(x)\rangle\rangle\right\rbrack\, ,
 \label{fq1}
\end{eqnarray}
where $p_h$  and $p$ are the four-momentum of the hadron and the
initial parton, respectively. After making all possible contraction
between the final constituent quark states with the thermal intermediate
states, one can obtain three distinctive contributions to the
above fragmentation function in medium,
 \begin{eqnarray}
   {\tilde D}_{q}^{h}(z_h, p^+)&=&{\tilde D}_q^{h(SS)}(z_h,p^+)
   +{\tilde D}_{q}^{h(ST)}(z_h, p^+) \nonumber \\
   &+&{\tilde D}_{q}^{h(TT)}(z_h, p^+)\, .
 \label{mesonF}
 \end{eqnarray}
The first contribution,
 \begin{eqnarray}
  {\tilde D}_{q}^{h(SS)}(z_h)
  &\approx& C_h\int_0^{z_h}\frac{dz_1}{2}
  R_h(0_\perp,\frac{z_1}{z_h}) \nonumber \\
 &\times& {\tilde F}_{q}^{q_1{\bar q}_2}(z_1, z_h-z_1,p^+)  \, ,
 \label{qfragm02}
 \end{eqnarray}
normally referred \cite{Hwa:2002tu} to as "shower-shower" quark
recombination comes from recombination of constituent quarks from
within the parton jet. It has exactly the same form as the parton
fragmentation functions in vacuum [Eq.(\ref{qfragm02})] in the
framework of quark recombination, except that the diquark
distribution function ${\tilde F}_{q}^{q_1{\bar q}_2}(z_1, z_2)$
are now also modified by the medium. Its definition is similar to
that in vacuum in Eq.~(\ref{qfrag03}) but the vacuum expectation
is replaced by thermal average. These modified diquark
distribution functions should in principle contain effects of
multiple scattering, induced gluon radiation and absorption, in
the same way as the modification of hadron fragmentation functions
in a thermal medium \cite{Osborne:2002dx}.

The second term in the modified fragmentation function,
 \begin{eqnarray}
  {\tilde D}_{q}^{h(ST)}(z_h, p^+)
  &=&\int_0^{z_h} \frac{dz_q}{z_h} \int\frac{d^2q_{\perp}}{2(2\pi)^3}
  \frac{R_h(q_\perp, z_q/z_h)}{(1-z_q/z_h)^2}\nonumber \\
& & \hspace{-1.3in}
\left [ f_q(q_{\perp},z_q p^+){\tilde F}_{q}^{\bar q}(z_h-z_q)
+f_{\bar q}(q_{\perp},z_q p^+){\tilde F}_{q}^{q}(z_h-z_q) \right ]
 \label{mesonST}
 \end{eqnarray}
is from recombination between a constituent quark (anti-quark)
from the parton jet and an anti-quark (quark) from the medium.
This is often referred to as ''thermal-shower'' quark
recombination. Here, $f_q$ and $f_{\bar q}$ are thermal quark
distributions, and ${\tilde F}_{q}^{q}(z)$ and ${\tilde
F}_{q}^{\bar q}(z)$ are single constituent quark or anti-quark
distributions of the fragmenting parton jet in a thermal medium
defined similarly as in the vacuum, except that the vacuum
expectation values are replaced again by thermal averaged
expectation. They should be different from the corresponding quark
distributions in vacuum because of multiple scattering, induced
gluon bremsstrahlung and parton absorption.

The final term in Eq.~(\ref{mesonF}),
 \begin{eqnarray}
   {\tilde D}_{q}^{h(TT)}(z_h, p^+)
   &=& Vp^+ \int\frac{d^2p_{h\perp}}{(2\pi)^3}
   \int_0^{z_h}\frac{dz_q}{2z_h}\int\frac{d^2q_{\perp}}{(2\pi)^3}
 \nonumber\\
   &&\hspace{-1.25in}
   f_q(q_{\perp}, z_{q}p^+)f_{\bar q}(p_{h\perp}-q_{\perp}, (z_h-z_q)p^+)
   R_h(q_{\perp}, \frac{z_q}{z_h})\,
 \label{mesonTT2}
 \end{eqnarray}
comes from recombination of two thermal constituent quarks. Here $V$ is the
total volume of the whole thermal system.
Since hadron production from the thermal quark recombination is not
correlated with the parton jet and its fragmentation, the above expression
is a little bit unnatural. One should be able to rewrite it in terms of
the invariant hadron spectrum from thermal quark recombination,
 \begin{eqnarray}
   (2\pi)^3\frac{dN^{h(TT)}}{dp_h^+d^2p_{h\perp}}
   & = &V\int_0^1 dx_q\int\frac{d^2q_{\perp}}{2(2\pi)^3}
   f_q(q_{\perp}, x_{q}p_h^+) \nonumber \\
& & \hspace{-1.0in}\times f_{\bar q}(p_{h\perp}-q_{\perp}, (1-x_q)p_h^+)
R_h(q_{\perp}, x_{q})\, .
 \label{mesonTT4}
 \end{eqnarray}
which is not correlated and therefore do not depend on the parton jet
fragmentation. The above expression also coincides with results from
other recombination models \cite{Fries:2003vb}.

\section{Conclusions}
\label{sec:4}

In this talk, I reviewed some new developments in the study of
modified single and dihadron fragmentation functions in dense
medium and their applications to heavy-ion collisions. All of them
will help to provide further test of the picture of parton energy
loss and jet quenching and enable more detailed characterization
of the dense medium those jets probe. The mass dependence of the
gluon formation time from the heavy quark leads to a unique change
of the medium size dependence of the heavy quark energy loss, from
linear to quadratic, when the initial quark energy and the
momentum scale are varied. The so-called ``dead-cone'' effect,
also caused by the heavy quark mass, in addition reduces the total
heavy quark energy loss. The form of heavy quark fragmentation
function into heavy quark mesons in vacuum, which is peaked at
large fractional momentum $z$, leads to a medium modification that
is different for light hadrons from massless partons. Since one
can identify heavy quark mesons, one can use them to tag heavy
quark propagation and study the difference between quark and gluon
energy loss \cite{Armesto:2005iq}. One can also use the energy
dependence of the suppression of single inclusive hadron spectra
to test the difference in quark and gluon energy loss due to the
non-Abelian gauge interaction. For fixed $p_T$ the fraction of
initial jets changes from quark to gluon dominated partons and the
different energy loss of quarks and gluons will result in a unique
energy dependence.

Finally, I also discussed the formulation of fragmentation
functions in a quark recombination picture within a constituent
quark model. Given the hadron's wavefunction in the constituent
quark model and neglecting interference effects, we have shown
that hadron fragmentation functions can be expressed as the
convolution of the recombination probability (given by the
hadron's wavefunction) and the constituent (or shower) quark
distribution of the jet. The constituent quark distributions are
defined as the overlapping matrices between parton fields and the
final constituent quark states, just like hadron fragmentation
functions as overlapping matrices between parton fields and final
hadrons. We have derived the DGLAP evolution equations for the
quark distribution functions. We then extended the formalism to
include the medium effect within the framework of field theory at
finite temperature. One naturally arrived at three distinctive
contributions from recombination between shower constituent
quarks, shower-thermal and thermal-thermal quarks, as have been
proposed by previous recombination models.

Since parton energy loss is only sensitive to the initial color
charge (or gluon) density of the medium, the observed jet quenching
points us to an enormously high initial gluon density created in
the central $Au+Au$ collisions at RHIC. However, parton energy loss
does not distinguish confined and deconfined matter. Quark recombination
between jet shower quarks and thermal quarks on the other hand is a
process of quark interaction that crosses hadronic boundary. If
proven, the combined signal of quark recombination
and thermalization will lead to unambiguous conclusion of deconfinement
of the produced dense matter.

I would like to thank Abhjit Majumder, Enke Wang and Beiwen Zhang
for their collaboration on the work I reported in this talk.
This work was supported the Director, Office of Energy
Research, Office of High Energy and Nuclear Physics, Division of
Nuclear Physics, and by the Office of Basic Energy Science,
Division of Nuclear Science, of  the U.S. Department of Energy
under Contract No. DE-AC03-76SF00098.


%

\end{document}